# Emergent kagome-like electronic texture in epitaxial β-Sb on Al(111)

Yang Wang[1, *], Pujing Zhang[2]
[1]Department of Physics, Hebei Normal University, Shijiazhuang 050024, China
[2]Department of Physics, Capital Normal University, Beijing 100048, China
[*]Authors to whom correspondence should be addressed: 790814014@qq.com

We report the epitaxial growth of β-Sb on Al(111) and the emergence of a kagome-like electronic texture with increasing Sb coverage. The surface evolves from disordered clusters through honeycomb and dice patterns before kagome-like regions appear. Diffraction, scanning tunneling microscopy, spectroscopy, and density functional calculations identify the ordered phase as β-Sb formed by AB stacking of two dice layers, with the honeycomb phase serving as a precursor. The kagome-like maxima occupy bridge positions within the same lateral lattice and differ in apparent height from dice regions by only about 25 pm. Together with the absence of an additional chemical component or distinct band manifold in photoemission, these observations support an electronic origin of the kagome-like contrast. Comparison of local tunneling spectra with calculated orbital projections suggests an energy selective change in the relative spectral contributions of Sb p states. Interface separation tests reveal sensitivity to Al–Sb coupling, while vacuum local density of states calculations identify states with enhanced bridge density. These results suggest that local interface changes could alter the balance between atomic and bridge contributions to tunneling within β-Sb. For a reference β-Sb model retaining the epitaxial geometry after Al removal, spin-orbit coupling modifies local band dispersions while the structure remains metallic. The model exhibits a finite intrinsic spin Hall response that varies with chemical potential.

## I. INTRODUCTION

Two-dimensional (2D) crystals are atomically thin solids with long-range order, following graphene's discovery, isostructural (graphene-like) 2D materials have been extensively explored [1-5]. Group-V elemental monolayers include black and blue phosphorene, arsenene, antimonene and bismuthene, among which antimonene is especially attractive owing to its tunable band gap [6-11]. First-principles calculations predict freestanding α- and β-antimonene (β-Sb) to be dynamically stable semiconductors [12], molecular beam epitaxy (MBE) has produced ordered 2D Sb phases on several metal substrates. Different growth pathways have been reported on metal substrates: on Ag(111) and Cu(111), α-Sb grows on $Ag_2Sb$ and $Cu_2Sb$ surface alloy layers, respectively; on Au(111), a surface alloy at submonolayer coverage evolves into a pure 2D Sb film at approximately one monolayer; and on W(110), α-Sb grows on a single layer of Sb atoms that passivates the substrate. [13-18]. Few layer β-Sb single crystals have also been grown on mica by van der Waals epitaxy, without detectable chemical bonding to the substrate [19]. These observations demonstrate the importance of distinguishing intrinsic Sb structures from substrate induced reconstructions and interfacial alloying. This distinction is particularly relevant because density functional theory (DFT) calculations predict that biaxial tensile strain and reduced buckling can drive freestanding β-Sb from a trivial semiconductor into a quantum spin Hall phase [20,21]. Recent calculations also have examined spin Hall responses in antimonene and related buckled monolayers. Equilibrium β-Sb was classified as a $Z_2$ trivial non-QSH phase through calculations of intrinsic spin Hall conductivity (SHC) and magnetic flux insertion [22], whereas α-Sb can host a $Z_2$ trivial phase with spin Chern number $C_s = 2$, in which spin Berry curvature is concentrated near Dirac points gapped by spin-orbit coupling (SOC) [23]. Epitaxial semimetallic α-Sb exhibits a finite, nonquantized spin Hall conductance that can be tuned by strain [24], while perpendicular electric fields modify the intrinsic SHC of silicene, germanene, stanene, and plumbene [25]. The kagome lattice, composed of corner sharing triangles, provides another platform for unconventional 2D electronic states. Simple kagome models support Dirac crossings and flat bands, while SOC can generate nontrivial Berry and spin Berry curvature [26-29]. Experimentally, a kagome-like germanene phase has been reported on Al(111) [30]. Beyond structural realizations, surface flat band states in FeSn kagome-honeycomb lateral heterostructures are suppressed near the interface and recover over orbital-dependent nanometer scale distances [31].

In this paper, we address the open question of whether an electronic kagome texture can emerge in antimonene. Here, we further demonstrate the phase evolution on Al(111) during MBE deposition. Angle-resolved photoemission spectroscopy (ARPES) of β-Sb resolves a Dirac-like crossing at Γ around −1.0 eV in experiment, consistent with DFT calculation, while a Wilson-loop (hybrid-Wannier) analysis of a Wannier90 tight-binding model yields $Z_2$ invariant $\nu = 0$, confirming its topologically trivial character. Calculations based on Wannier functions further show that the metallic bands of β-Sb generate pronounced positive and negative contributions to the spin Berry curvature near K and K′ and along Γ-M. Integration over the Brillouin zone yields an intrinsic spin Hall conductivity that varies substantially with the Fermi energy. Together, these results establish epitaxial β-Sb on Al(111) as a platform combining an electronic texture resembling a kagome lattice with a tunable intrinsic spin Hall response.

## II. EXPERIMENTAL AND COMPUTATIONAL METHODS

### A. Sample preparation and structural characterization

The bulk crystal structure of aluminum is face-centered cubic (fcc) with a lattice constant of 4.04 Å, and Al(111) surface exhibits hexagonal symmetry with 1×1 reconstruction [32]. The substrate is cleaned through cycles of $Ar^+$ sputtering and annealing at 550 °C. The resulting flat and wide terraces are imaged by scanning tunneling microscope (STM) at 77 K under an ultra-high vacuum of $1\times10^{-10}$ mbar, as shown in Fig. 1(a). The line profile in Fig. 1(b) confirms monoatomic step heights and the absence of residual islands or vacancies, indicating an atomically ordered and chemically clean surface. This uniform adsorption landscape enforces a unique epitaxial registry for the antimonene overlayer, thereby enabling large area and single crystal growth.

Consequently, sharp reflection high-energy electron diffraction (RHEED) streaks that can furnish insights into the reciprocal lattice are shown in Fig. 1(c) and Fig. 1(d). Those two patterns correspond to incidence azimuths that differ by 30°, indicating that the $[2\bar{1}\bar{1}]$ and $[1\bar{1}0]$ crystal orientations give rise to wide and narrow stripes, respectively. After substrate preparation, Sb atoms are evaporated from an effusion cell held at 350 °C, while the growth chamber of the MBE system is maintained at room temperature under $1\times10^{-10}$ mbar. Throughout MBE growth, only one distinct new set of diffraction streaks in Fig. 1(e) and Fig. 1(f) is observed, exhibiting a spacing exactly one third of bare Al(111), while preserving the same sixfold rotational symmetry. This demonstrates that the epitaxial Sb overlayer is azimuthally locked to Al(111) and its lattice constant is three times that of Al(111) along $\langle 2\bar{1}\bar{1}\rangle$ and $\langle 1\bar{1}0\rangle$ crystallographic direction families. The new streaks persist during subsequent Sb deposition and post deposition annealing up to the substrate cleaning setpoint, attesting to the thermal robustness of the alloy free Al(111) template and the epitaxial locking of the Sb overlayers.

Samples with different nominal Sb coverages were prepared by varying the deposition time up to 8 min, and their surface morphologies were characterized by STM. Figs 2(a)–2(d) show representative images of the initial disordered surface and three distinct periodic surface patterns observed at different deposition times. With increasing deposition time, the ordered regions display a hexagonal network, a similar network with additional bright protrusions at the hexagon centers, and triangular motifs connected at their corners around hexagonal dark regions [33]. These STM patterns are referred to as honeycomb, dice, and kagome-like, respectively.

### B. Electronic structure measurements and spectroscopic analysis

Three sets of local differential conductance *dI*/*dV* spectra were acquired at 77 K at selected positions within three STM patterns, as indicated by the colored markers in Fig. 2(e). During the scanning tunneling spectroscopy (STS) measurements, the tip height was held fixed while the sample bias was swept from −2.0 to +2.0 V. Forward and backward sweeps were recorded with 400 bias points in each direction. The differential conductance was measured using lock-in detection, with settling and integration times of 60 ms each at every bias point.

We tracked the compositional evolution during deposition by X-ray photoelectron spectroscopy (XPS), acquired using an Al $K_\alpha$ excitation source. The SOC splits the Sb 4*d* core level into $4d_{5/2}$ and $4d_{3/2}$ with an area ratio of 3:2 [34-36]. Spectral fitting included a background contribution, and the peak positions and line shapes were compared across the deposition series, as shown in Fig. 2(f). The Sb 4*d* doublet shifted by approximately 0.15 eV toward lower binding energies between 1 and 4 min. From 4 to 10 min, the two peaks remained near 31.70 and 32.95 eV, while the Sb 4d intensity continued to increase. The Al 2*p* peak remained near 73.0 eV throughout the deposition series and retained a similar overall line shape. To experimentally resolve the energy dispersion along the M-Γ-M direction, ARPES measurements were performed using a He I light source at 7 K and a base pressure of $3\times10^{-11}$ mbar. The ARPES spectrum in Fig. 3(a) remains essentially unchanged for deposition times below 8 min. From the *E*-*k* dispersion extracted using the 2D curvature method in Igor [37], a Dirac like crossing is resolved at the Γ point at approximately −1.2 eV below the Fermi level, as shown in Fig. 4(b).

To compare the energy dependent spectral features of the dice and kagome-like regions, residual spectra were obtained by subtracting a slowly varying background from the normalized *dI*/*dV* spectra. The forward and backward sweeps were ordered by increasing sample bias on a common grid containing 400 points over the range from −2.0 to +2.0 V. Spectra exhibiting pronounced forward/backward disagreement or junction instability were excluded. Each retained sweep was normalized by subtracting its median signal and dividing by the difference between its 95th and 5th percentiles, with these quantities evaluated over the bias interval from −1.9 to +1.9 V.

The normalized sweeps were smoothed using Savitzky–Golay filtering [38] with polynomial order 3 and a window of 17 points, then averaged within each spatial group. The slowly varying background was estimated using the same method with a window of 101 points and subtracted from each mean spectrum to obtain the residual. The final dice residual represents six spatial positions from two STM fields acquired after 6 min of deposition, with three positions in each field. The kagome-like residual represents five spatial positions from an STM field acquired after 8 min. Group residuals were combined with weights proportional to their numbers of spatial positions, giving equal weight to each retained position; both sweep directions contributed equally. The resulting residuals quantify relative departures from the selected smooth background in arbitrary units.

To provide an orbital-resolved theoretical reference for these residual spectra, projected densities of states were extracted from the VASP DOSCAR output for the relaxed 2 × 2 AB-stacked β-Sb/Al(111) model, the final Fermi energy recorded in OUTCAR was taken as 0 eV. Atomic projections were summed separately for the Al substrate and the Sb film. The Sb atoms were further grouped according to their heights along the surface normal. The highest Sb height group was identified as the uppermost Sb sublayer, while the remaining Sb atoms were combined into a lower Sb group. The Sb *p* PDOS was separated into the summed contributions from the two orbitals parallel to the surface and the contribution from the orbital normal to it. These components and the total Sb *p* PDOS are presented in Fig. 4(h), together with the mean STS residuals of the dice and kagome-like regions.

### C. First-principles calculations

Calculations were performed within density functional theory using the projector augmented wave method and the exchange and correlation functional of Perdew, Burke, and Ernzerhof (PBE) within the generalized gradient approximation (GGA), as implemented in the Vienna Ab Initio Simulation Package (VASP) [39-43]. Structural relaxations employed an energy cutoff of 500 eV for the plane wave basis, Gaussian smearing, and a 7×7×1 sampling mesh centered at Γ. The electronic convergence threshold was $1\times10^{-5}$ eV, and the residual force on each atom was reduced below 0.02 eV/$Å^{-1}$. For the band calculations, all Al atoms were removed from the relaxed Sb/Al(111) slab, while the optimized geometry of the remaining Sb layer was retained without further relaxation.

After obtaining a self-consistent charge density for this isolated β-Sb model, the band energies were evaluated without updating the charge density along K-Γ-M-K, using 90 points per segment. Calculations were performed with and without SOC, using an energy cutoff of 500 eV, Gaussian smearing of 0.05 eV, and an electronic convergence criterion of $1\times10^{-6}$ eV. SOC was treated within the noncollinear formalism with the spin quantization axis along z, and symmetry was disabled in both calculations. The resulting band structures were processed using VASPKIT [44, 45]and are shown in Fig. 5(a).

STM simulations [46] were performed for the optimized 2×2 Sb/Al(111) model containing 108 Al and 24 Sb atoms. A partial charge density containing the unoccupied states from the Fermi level to 0.50 eV above it was generated from the converged wave functions using a 6×6×1 mesh and 500 eV cutoff. Within the Tersoff–Hamann approximation [47], the partial charge density sampled at a constant height was interpreted as a relative tunneling current. The map used for comparison with experiment was calculated at 3.5 Å over a 10×10 $nm^2$ area on a 512×512 grid. A periodic in-plane Gaussian filter with a standard deviation of 0.60 Å was applied to represent lateral tip broadening.

For quantitative comparison, matching 5×5 $nm^2$ regions with 256×256 pixels were selected from the experimental backward current image and the simulated relative current map. The simulated map calculated at a height of 3.5 Å was rotated clockwise by 10° to match the experimental lattice orientation. The simulated STM current map is shown in Fig. 4(b) for comparison with the experimental STM current image. The magnitudes of the 2D Fourier transforms were logarithmically scaled and normalized independently for comparison. Strip averaged profiles were sampled at 840 equally spaced points along a 4.2 nm direction perpendicular to the dominant rows, while the signal was averaged over a width of 3.0 nm parallel to the rows. A slowly varying background obtained using a Gaussian filter with a standard deviation of 0.70 nm was subtracted from each profile, after which the resulting signals were independently scaled between their first and 99th percentiles. For the plotted comparison, the profiles were lightly smoothed using a Gaussian kernel with a standard deviation of 1.5 sampling intervals, and the simulated profile was shifted by −0.160 nm along the distance axis to align its phase with the experimental profile.

### D. Wannier interpolation for topological and intrinsic spin Hall conductivity

For Wannier interpolation [48], the substrate was removed from the relaxed Sb/Al(111) model while retaining the optimized β-Sb structure. A fixed charge density calculation including SOC was then performed using a Γ-centered 12×12×1 mesh, in a plane wave cutoff of 500 eV, and Gaussian smearing of 0.05 eV. Crystal symmetry was disabled during this calculation. The Bloch eigenvalues, overlap matrices, and projection matrices were generated through the VASP-Wannier90 interface [49]. A total of 72 spinor Bloch bands were mapped onto 18 spinor Wannier functions initialized from Sb *p* orbitals. An outer disentanglement window spanning approximately ±2.50 eV about the Fermi level was used.

Using a mixing ratio of 0.7 and a convergence tolerance of $10^{-8}$, the disentanglement procedure converged after 386 iterations. The resulting Wannier Hamiltonian in real space was used for band interpolation and the subsequent intrinsic spin Hall conductivity calculations. The interpolated band structure was evaluated along K-Γ-M-K using 90 k points for each path segment.

The intrinsic spin Hall conductivity was evaluated from the Wannier Hamiltonian using WannierBerri [50] and the Qiao formulation [51]. The required spin matrix elements were read from the Wannier90 spin matrix file. The static xy component with spin polarization along z was calculated on a uniform 301×301×1 mesh at zero frequency. The SHC was evaluated at 121 equally spaced chemical potentials within a ±0.30 eV window centered on the Fermi level, giving an energy spacing of 5 meV. A thermal energy of 0.026 eV was used in the Fermi distribution. The conductivity obtained in S $m^{-1}$ was converted to a 2D sheet conductivity by multiplying it by the out-of-plane lattice parameter of the simulation cell, and was reported in μS. The calculated intrinsic SHC as a function of chemical potential is shown in Fig. 5(b).

To identify the momentum-space origin of the spin Hall response, the spin Berry curvature was evaluated as the *k*-resolved integrand of the static spin Hall conductivity within the Qiao formulation. The calculation was performed at the Fermi level on a uniform 301×301×1 mesh covering the full Brillouin zone, without symmetry reduction or additional *k* space refinement. The component corresponding to a spin current flowing along x, an electric field applied along y, and spin polarization along z was extracted. The resulting spin Berry curvature distribution in momentum space is shown in Fig. 5(c).

Because the Wannier interpolated spectrum has no global insulating gap, the topology was evaluated for a fixed, directly isolated band subspace rather than for bands selected by the physical Fermi level. The lowest 12 bands of the 18-band spinor Wannier Hamiltonian were chosen, corresponding to the boundary between bands 30 and 31 in the original DFT calculation. Their direct separation was examined throughout the 2D Brillouin zone on a uniform 301×301×1 k point mesh. The region surrounding the minimum gap was refined twice using local 101×101 meshes, with $10^{-5}$ eV adopted as the numerical threshold for gap closure. After confirming that the selected subspace remained directly separated, its $Z_2$ invariant was calculated using Z2Pack [52] from the evolution of the Wannier charge centers over half of the Brillouin zone. The Wilson loops were closed along $k_1$ and swept from $k_2 = 0$ to $k_2 = 1/2$ at $k_3 = 0$. The calculation used 101 initial lines and allowed 30 to 300 k points along each loop in increments of 10. The calculation yielded $Z_2$= 0 for the selected 12-band subspace.

## III. RESULTS AND DISCUSSION

### A. Epitaxial growth and evolution of Sb with increasing coverage

The STM sequence in Fig. 2 establishes a systematic relation between deposition time and the observed surface patterns. With the Sb flux and substrate temperature held constant during growth, deposition time was used as a relative measure of nominal Sb coverage. The STM images in Figs. 2(a)–2(d) show the evolution of the surface morphology of deposition. After 2 min, the surface is dominated by irregular protrusions and small clusters, with no extended periodic order. Ordered regions displaying a honeycomb pattern appear at 4 min alongside the remaining disordered areas. Additional bright protrusions become visible near the centers of the hexagonal units at 6 min, producing a dice pattern. In the 8 min image, spatially distinct regions with dice and kagome-like contrast are observed on the ordered surface. These images document the successive appearance of honeycomb, dice, and kagome-like STM patterns with increasing deposition time.

The evolution of the STM patterns is accompanied by distinct changes in the tunneling spectra, as shown in Fig. 2(e). At 4 min, spectra acquired at three positions in the honeycomb region exhibit a smooth, broadly U-shaped bias dependence, with reduced conductance around zero bias. At 6 min, more pronounced spectral features emerge, including occupied state peaks near −1.1 and −0.5 V and an unoccupied state peak near +0.8 V. At 8 min, the spectra acquired in the regions with kagome-like contrast are dominated by an occupied state feature near −1.2 V, while the positive bias feature is less prominent relative to this peak. The spectra within each region retain similar overall shapes, although local variations in peak intensity are visible, particularly at 8 min. These changes in spectral shape and relative peak prominence are consistent with an evolution of the local electronic states probed by tunneling as growth proceeds.

Fig. 2(f) shows the corresponding evolution of the core-level spectra. The Sb 4*d* intensity increases with deposition time, consistent with progressive Sb accumulation. The fitted Sb 4*d* peaks shift toward lower binding energies by approximately 0.15 eV between 1 and 4 min, after which their positions remain near 31.70 and 32.95 eV. The Al 2*p* peak remains close to 73.0 eV throughout the deposition series. No distinct additional core-level component is resolved as the dice and kagome-like patterns emerge. The continued evolution of the tunneling spectra after the Sb 4*d* peak positions have largely stabilized therefore supports growth-associated changes in the local electronic structure, without a concurrent appearance of a resolvable new chemical component in XPS.

## B. Structural identification of the STM patterns

We identify the ordered Sb film exhibiting the dice pattern as β-Sb formed by AB stacking of two dice layers. In the structural model shown in Fig. 4(c), the two layers have the same orientation, while the upper layer is displaced laterally by $(a_1 + a_2)/3$ relative to the lower layer. This registry places the centers of the upper layer hexagonal units above threefold hollow sites in the lower layer. In the relaxed configuration, the combined atomic arrangement forms the buckled honeycomb framework of β-Sb, providing a structural description of the film underlying the observed dice contrast.

This assignment is consistent with the lateral periodicity established by RHEED and STM. Compared with the bare Al(111) patterns in Figs. 1(c) and 1(d), the additional streaks in Figs. 1(e) and 1(f) have approximately one-third the spacing along the corresponding azimuths. This reciprocal space relation indicates a surface period approximately three times that of Al(111). The line profile in Fig. 4(a) gives a lateral period of approximately 8.48 Å. The agreement between these real space and reciprocal space observations constrains the periodicity and epitaxial orientation of the proposed Sb structure.

Structural relaxation in DFT further favors the AB stacked model within the examined set of candidates. Only this configuration yielded a converged relaxed structure, whereas the honeycomb configuration and the other tested arrangements did not yield converged relaxed models. The experimental STM current image and the simulated relative current map in Fig. 4(b) exhibit periodic row contrast with comparable lateral spacings. The simulation was obtained from the relaxed Sb/Al(111) structure at a fixed height of 3.5 Å. After orientational alignment, the FFT insets show corresponding principal peak groups with approximately hexagonal arrangements and similar reciprocal space separations.

The simulated contrast was compared with the experimental backward current image of the 6 min dice region shown in Fig. 4(b), acquired at a sample bias of +0.5 V and a current setpoint of approximately 600 pA. The simulated relative current map was evaluated using unoccupied states from 0 to +0.50 eV relative to the Fermi level, corresponding to the experimental bias window. To compare their periodic spatial modulation, strip averaged profiles were extracted from matching $5 \times 5$ nm$^2$ regions with $256 \times 256$ pixels. The signal was averaged over a 3.0 nm width parallel to the dominant rows and sampled along a 4.2 nm direction perpendicular to them. Following background subtraction and independent normalization, the simulated profile was shifted along the distance axis to align its phase with the experimental profile. The curves in Fig. 4(d) show closely matching periodic modulations, with dominant periods of approximately 0.42 and 0.43 nm for experiment and simulation, respectively. Their corresponding doubled periods of approximately 0.85 and 0.86 nm agree with the surface periodicity established by STM and RHEED. The agreement in reciprocal space peak geometry and normalized real space modulation further supports the AB stacked β-Sb structural model.

The measured band dispersion in Fig. 3(b) and the calculated band structure in Fig. 5(a) both exhibit a Γ-centered Dirac-like feature at binding energy range around −1 eV relative to the Fermi level. This qualitative correspondence in momentum space location and dispersion provides complementary electronic support for the proposed AB stacked β-Sb framework. The latter were obtained after removing Al from the relaxed slab while retaining the optimized Sb geometry. The correspondence in momentum space location and general dispersion is qualitative, with a difference in binding energy between experiment and calculation. It nevertheless provides additional support for the proposed β-Sb framework alongside the diffraction, relaxation, and STM results.

The kagome-like contrast emerging at higher Sb coverage is assigned to an electronic texture of the AB stacked β-Sb film rather than to an additional kagome atomic layer. The kagome-like electronic texture is observed in STM images acquired at sample biases of −0.30, −0.50, −0.60, and +0.60 V. Its observation at both negative and positive sample bias demonstrates that the pattern is accessible when probing either occupied or unoccupied states. The schematic registry in Fig. 4(g) maps the kagome-like intensity maxima onto bridge positions of the projected dice framework, forming a network within the existing surface unit cell. The line profiles in Figs. 4(a) and 4(e) give the same characteristic periodicity of approximately 8.48 Å. Their FFT insets also exhibit the same approximately sixfold arrangement of the principal peaks, with matching reciprocal-space separations and orientations. Thus, the appearance of kagome-like contrast with increasing deposition time preserves the lateral periodicity and orientation established during the preceding growth stages.

The coexistence of dice and kagome-like regions in Fig. 4(f) allows their apparent heights to be compared under identical tunneling conditions. The profile across the boundary shows an apparent height difference of only approximately 25 pm. The small offset supports continuity of the β-Sb surface across the two regions and is consistent with a change in electronic contrast. Because the measured STM height includes contributions from the local electronic states, the offset should not be interpreted as a geometric interlayer separation.

The spectroscopic measurements provide complementary evidence for the persistence of the underlying film. As shown in Fig. 2(f), the Sb 4*d* doublet exhibits no appreciable binding energy shift or newly resolved chemically shifted component accompanying the appearance of kagome-like contrast. The ARPES measurements likewise retain a single resolved set of dispersive bands at a deposition time of 8 min, when kagome-like regions are already observed by STM. No additional resolved band manifold appears at this stage. These spatially averaged measurements support continuity of the dominant β-Sb film.

The comparison in Fig. 4(h) provides further electronic support for identifying the dice regions as β-Sb formed by AB stacking of two dice layers. The dice STS residual exhibits two occupied state peaks, $P_1$ and $P_2$, near −1.10 and −0.55 eV, respectively. These correspond qualitatively to features in the calculated Sb *p* PDOS near −1.21 and −0.58 eV. The calculated feature associated with $P_1$ is dominated by the p orbital normal to the surface, whereas that associated with $P_2$ has a larger contribution from the two p orbitals parallel to the surface. On the unoccupied side, $P_3$ near +1.34 eV corresponds to the calculated feature near +1.42 eV, while $P_4$ near +1.86 eV occurs within the higher energy group of PDOS features spanning approximately +1.7 to +2.0 eV. The calculated states in this range are dominated by the *p* orbitals parallel to the surface. This correspondence across both occupied and unoccupied states complements the agreement in lattice periodicity, simulated STM contrast, and band dispersion, supporting the assignment of the dice regions to the β-Sb structure with AB stacking.

The STS residual changes in Fig. 4(h) can be interpreted qualitatively through comparison with the calculated PDOS for the upper Sb sublayer. The enhanced kagome-like maximum near −1.20 eV coincides with a feature dominated by *p* orbitals normal to the surface, while the maximum near −1.63 eV lies in a region containing mixed normal and parallel *p* contributions. The positive dice features around $P_2$, $P_3$, and $P_4$, associated with larger calculated contributions from p orbitals parallel to the surface, weaken in the kagome-like residual. The changes therefore indicate an energy selective reorganization of the orbital contributions expressed in the local tunneling spectrum within the established β-Sb framework.

### C. Mechanistic insights into the kagome-like electronic texture

The structural analysis identifies the kagome-like contrast as an electronic texture within the β-Sb framework formed by AB stacking of two dice layers. Its systematic emergence at greater Sb coverage motivates examining whether changes in the local interface environment can redistribute the states contributing to tunneling. We investigated this possibility using an Al–Sb separation test, orbital analysis of the experimental STS differences, and reconstruction of the vacuum LDOS.

The separation test used a reference cell containing 27 Al and 6 Sb atoms. The Sb film was displaced rigidly by 0.20 and 0.40 Å away from Al(111), while the Al coordinates, Sb internal geometry, lateral registry, and cell dimensions were retained. For each separation, the electronic density was converged with SOC on a 6 × 6 × 1 k mesh, with no further ionic relaxation. Orbital PDOS were integrated over selected energy windows for the two Sb height groups and the uppermost Al plane, and the trends were checked against sums of discrete eigenstate projections. This construction isolates the electronic response to a change in interface separation within the existing structural framework.

Increasing the Al–Sb separation mainly affected the Sb *p* states normal to the surface. Their contribution rose markedly in the unoccupied window, while the parallel *p* component changed little. The occupied response varied with energy; near the Fermi level, the normal component decreased slightly and the parallel component increased slightly. In the same occupied window, Sb *p* weight increased in the upper group and decreased in the lower group, together with a reduction in the surface Al *s* and *p* contribution. The total Sb *p* weight changed little. These results show that interface coupling influences both the orbital composition of Sb states and their distribution between the two height groups.

The STS residual changes in Fig. 4(h) can be interpreted qualitatively through comparison with the calculated PDOS for the upper Sb sublayer. The enhanced kagome-like maximum near −1.20 eV coincides with a feature dominated by *p* orbitals normal to the surface, while the maximum near −1.63 eV lies in a region containing mixed normal and parallel *p* contributions. The positive dice features around $P_2$, $P_3$, and $P_4$, associated with larger calculated contributions from *p* orbitals parallel to the surface, weaken in the kagome-like residual.

This correspondence suggests that the emergence of kagome-like contrast accompanies changes in the relative spectral contributions of Sb *p* states with different orbital character. The normal *p* related feature becomes more prominent near −1.20 eV, whereas selected parallel *p* related features lose prominence. The changes therefore indicate an energy selective reorganization of the orbital contributions expressed in the local tunneling spectrum within the established β-Sb framework.

To examine this interpretation further, we subtracted the mean dice residual from the mean kagome-like residual and compared the difference with orbital response templates constructed from the calculated PDOS. These represented changes in orbital weight, small energy shifts, and redistribution between the two Sb height groups. The same smoothing and background subtraction were applied to experiment and theory. The difference showed a modest correspondence with redistribution of normal p weight between the Sb groups in the deeper occupied spectrum, while the correspondence weakened closer to the Fermi level.

The STM pattern also depends on how these states extend into vacuum. We therefore reconstructed the vacuum LDOS of the reference structure from its SOC wave functions and compared the densities above Sb sites and bridge positions. A group of states near Γ showed bridge enhancement in all three connection directions, with contributions from upper Sb p orbitals parallel to the surface and from surface Al s and p states. Across the screened states, stronger bridge tendencies generally accompanied a larger parallel p fraction, whereas the separation test mainly affected normal p states.

This group lies below the principal negative bias imaging windows and contributes very little to their integrated density. When all sampled states are summed over the complete occupied and unoccupied imaging windows, the density above the upper Sb sites remains greater than that above the bridges. Under the tested conditions, the ideal reference structure therefore does not reproduce the observed bridge contrast. The mechanism must account for a change in these relative spatial contributions within the full imaging window.

A possible mechanism is that local changes at the Al–Sb interface during growth alter orbital mixing and the balance between atomic and bridge contributions to tunneling. Such changes could shift the energies of relevant states and modify their extension and interference in vacuum. A weaker atomic contribution, a stronger bridge contribution, or both could then make the kagome-like network visible within the existing β-Sb framework.

The specific change induced by growth remains unresolved. Variations in registry, buckling, strain, defects, or interface potential could provide the perturbation. The calculations have not yet reproduced the contrast reversal or established how the normal p response couples to the bridge contributions.

### D. Electronic structure and intrinsic spin Hall response

The electronic structure and intrinsic spin Hall response were calculated for the β-Sb framework after removing Al from the relaxed slab while retaining the Sb geometry. Fig. 5(a) compares the band structures with and without SOC along K–Γ–M–K. The calculated dispersion is qualitatively consistent with the ARPES bands in Fig. 3(b), including the Γ-centered Dirac-like feature near −1.0 eV. SOC modifies the band separations and dispersions near several crossings, while the overall dispersion remains similar. Bands continue to cross the Fermi level, indicating that the local changes induced by SOC leave the reference β-Sb structure metallic.

The corresponding 2D sheet SHC is shown in Fig. 5(b). It is finite at the reference Fermi level and remains positive throughout the chemical potential interval from −0.30 to +0.30 eV. As the chemical potential rises, the SHC increases to a broad maximum of approximately 3 μS near +0.12 eV and then decreases gradually. Lowering the chemical potential suppresses the response more strongly than the enhancement obtained by raising it. This asymmetry indicates that states on the two sides of the Fermi level contribute differently to spin transport, making the intrinsic response sensitive to band filling.

Fig. 5(c) shows the spin Berry curvature summed over bands with occupations determined by the reference Fermi level. The distribution is weak around Γ and more pronounced towards the outer parts of the Brillouin zone, including regions near M. Positive and negative patches indicate competing contributions from different momenta, with the finite SHC reflecting their net contribution across the Brillouin zone. Changing which states are occupied alters this balance, providing a microscopic basis for the chemical potential dependence in Fig. 5(b). The response therefore depends on both the momentum distribution of spin Berry curvature and the filling of the contributing bands.

The selected band subspace has a trivial $Z_2$ classification. Together with the metallic bands and finite SHC in Fig. 5, this result characterizes the reference β-Sb model as a metal supporting a nonquantized intrinsic spin Hall response. This response arises from the net spin Berry curvature of the occupied states, and its dependence on chemical potential reflects changes in band filling.

## IV. CONCLUSIONS

Combined structural and spectroscopic measurements, supported by density functional calculations, identify the ordered Sb film on Al(111) as β-Sb formed by AB stacking of two dice layers. With increasing coverage, the surface evolves from disordered clusters through a honeycomb precursor to dice domains, followed by kagome-like regions. The kagome-like maxima occupy bridge positions within the same lateral lattice, with an apparent height difference of only about 25 pm from dice regions. The common periodicity and orientation, together with the absence of a resolved new chemical component or separate band manifold in photoemission, support an electronic texture within the β-Sb framework.

The STS residuals of the dice and kagome-like regions, interpreted through calculated orbital projections, suggest an energy selective change in the relative spectral contributions of Sb *p* states. Interface separation tests show that Al–Sb coupling selectively modifies orbital contributions and their distribution between Sb sublayers. Vacuum LDOS reconstruction identifies bridge enhancement in selected states containing contributions from Sb p orbitals parallel to the surface and surface Al states. These findings suggest that local interface changes during growth could alter orbital mixing and the balance between atomic and bridge contributions to tunneling. Atomic contributions remain stronger in the ideal model over the tested imaging windows, leaving the specific perturbation responsible for the observed bridge contrast unresolved.

For a β-Sb reference model retaining the epitaxial geometry after Al removal, spin–orbit coupling modifies local band dispersions while metallicity persists. The examined band subspace has a trivial $Z_2$ classification. The model exhibits a finite, nonquantized intrinsic spin Hall response reflecting the balance between positive and negative spin Berry curvature contributions across the Brillouin zone. This response varies with chemical potential as band filling changes.

**Reference**

[1] K. S. Novoselov, A. K. Geim, S. V. Morozov, D. Jiang, Y. Zhang, S. V. Dubonos, I. V. Grigorieva, and A. A. Firsov, Science **306**, 1102896 (2004).

[2] K. S. Novoselov, A. K. Geim, S. V. Morozov, D. Jiang, M. I. Katsnelson, I. V. Grigorieva, S. V. Dubonos, and A. A. Firsov, Nature **438**, 04233 (2005).

[3] C. L. Kane and E. J. Mele, Phys. Rev. Lett. **95**, 226801 (2005).

[4] J.-Y. Wang, S.-B. Deng, Z.-F. Liu, and Z.-R. Liu, Natl. Sci. Rev. **2**, 080 (2015).

[5] A. Molle, J. Yuhara, Y. Yamada-Takamura, and Z. Sofer, Chem. Soc. Rev. **54**, 1845 (2025).

[6] S.-L. Zhang, M.-Q. Xie, F.-Y. Li, Z. Yan, Y.-F. Li, E.-J. Kan, W. Liu, Z.-F. Chen, and H.-B. Zeng, Angew. Chem. Int. Ed. **55**, 1666 (2016).

[7] L.-K. Li, Y.-J. Yu, G.-J. Ye, Q.-Q. Ge, X.-D. Ou, H. Wu, D.-L. Feng, X.-L. Chen, and Y.-B. Zhang, Nat. Nanotechnol. **9**, 372 (2014).

[8] J.-L. Zhang, S.-T. Zhao, C. Han, Z.-Z. Wang, S. Zhong, S. Sun, R. Guo, X. Zhou, C.-D. Gu, K.-D. Yuan, Z.-Y. Li, and W. Chen, Nano Lett. **16**, 4903 (2016).

[9] S.-L. Zhang, Z. Yan, Y.-F. Li, Z.-F. Chen, and H.-B. Zeng, Angew. Chem. Int. Ed. **54**, 3112 (2015).

[10] J. A. Carrasco, P. Congost-Escoin, M. Assebban, and G. Abellán, Chem. Soc. Rev. **52**, 1288 (2023).

[11] F. Reis, G. Li, L. Dudy, M. Bauernfeind, S. Glass, W. Hanke, R. Thomale, J. Schäfer, and R. Claessen, Science **357**, 287 (2017).

[12] G.-X. Wang, R. Pandey, and S. P. Karna, ACS Appl. Mater. Interfaces **7**, 11490 (2015).

[13] J.-P. Ji, X.-F. Song, J.-Z. Liu, Z. Yan, C.-X. Huo, S.-L. Zhang, M. Su, L. Liao, W.-H. Wang, Z.-H. Ni, Y.-F. Hao, and H.-B. Zeng, Nat. Commun. **7**, 13352 (2016).

[14] Y. Shao, Z.-L. Liu, C. Cheng, X. Wu, H. Liu, C. Liu, J.-O. Wang, S.-Y. Zhu, Y.-Q. Wang, D.-X. Shi, K.

Ibrahim, J. T. Sun, Y. L. Wang, and H. J. Gao, Nano Lett. **18**, 2133 (2018).
[15] S.-Y. Zhu, Y. Shao, E. Wang, L. Cao, X.-Y. Li, L.-Z. Liu, C. Liu, L.-W. Liu, J.-O. Wang, O. Ibrahim, J.-T. Sun, Y.-L. Wang, S.-X. Du, and H.-J. Gao, Nano Lett. **19**, 6323 (2019).
[16] E. D. Cantero, E. A. Martínez, L. N. Serkovic-Loli, J. D. Fuhr, O. Grizzi, and E. A. Sánchez, J. Phys. Chem. C **125**, 9273 (2021).
[17] S. Sun, T. Yang, Y.-Z. Luo, J. Gou, Y.-L. Huang, C.-D. Gu, Z.-R. Ma, X. Lian, S.-S. Duan, A. T. S. Wee, M. Lai and J. L. Zhang, J. Phys. Chem. Lett. **11**, 8976 (2020).
[18] P. Zhang, C. Ma, S.-X. Sheng, H. Liu, J.-S. Gao, Z.-J. Liu, P. Cheng, B.-J. Feng, L. Chen, and K.-H. Wu, Phys. Rev. Mater. **6**, 074002 (2022).
[19] T. Jaroch, L. Żurawek-Wyczesany, A. Stępniak-Dybala, M. Krawiec, M. Kopciuszyński, P. Dróżdż, and M. Gołębiowski, Nano Lett. **24**, 12469 (2024).
[20] S. K. Radha and W. R. L. Lambrecht, Phys. Rev. B **101**, 235111 (2020).
[21] M.-W. Zhao, X.-M. Zhang, and L.-Y. Li, Sci. Rep. **5**, 16108 (2015).
[22] A. C. Tyner and P. Goswami, Sci. Rep. **13**, 11393 (2023).
[23] B. Wang, X. Zhou, Y.-C. Hung, Y.-C. Lin, H. Lin, and A. Bansil, 2D Mater. **11**, 025033 (2024).
[24] J. Cook, P.-Y. Yang, T. Volz, C. Conner, R. Satterfield, J. Berglund, Q. Lu, R. G. Moore, Y.-T. Yao, T.-R. Chang, and G. Bian, Mater. Today Phys. **51**, 101664 (2025).
[25] S. M. Farzaneh and S. Rakheja, Phys. Rev. B **104**, 115205 (2021).
[26] L. Lei, J. Dai, H. Dong, Y. Geng, F. Cao, C. Wang, R. Xu, F. Pang, Z.-X. Liu, F. Li, Z. Cheng, G. Wang, and W. Ji, Nat. Commun. **14**, 6320 (2023).
[27] Z.-M. Zhang, B.-C. Gong, J.-H. Nie, F. Meng, Q. Zhang, L. Gu, K. Liu, Z.-Y. Lu, Y.-S. Fu, and W. Zhang, Nano Lett. **23**, 954–961 (2023).
[28] C. Chen, J. Zheng, R. Yu, S. Sankar, K. T. Law, H. C. Po, and B. Jäck, Phys. Rev. Research **5**, 043269 (2023).
[29] D. Di Sante, C. Bigi, P. Eck, S. Enzner, A. Consiglio, G. Pokharel, P. Carrara, P. Orgiani, V. Polewczyk, J. Fujii, P. D. C. King, I. Vobornik, G. Rossi, I. Zeljkovic, S. D. Wilson, R. Thomale, G. Sangiovanni, G. Panaccione, and F. Mazzola, Nat. Phys. **19**, 1135–1142 (2023).
[30] O. Kubo, S. Kinoshita, H. Sato, K. Miyamoto, R. Sugahara, S. Endo, H. Tabata, T. Okuda, and M. Katayama, Phys. Rev. B **104**, 085404 (2021).
[31] T. A. Pham, S.-H. Kang, Y. Ozbek, M. Yoon, and P. Zhang, ACS Nano **18**, 8768–8776 (2024).
[32] J. Wintterlin, J. Wiechers, H. Brune, T. Gritsch, H. Höfer, and R. J. Behm, Phys. Rev. Lett. **62**, 59 (1989).
[33] R. Kotecký, J. Salas, and A. D. Sokal, Phys. Rev. Lett. **101**, 030601 (2008).
[34] J. Sinova, S. O. Valenzuela, J. Wunderlich, C. H. Back, and T. Jungwirth, Rev. Mod. Phys. **87**, 1213 (2015).
[35] E. McCafferty and J. P. Wightman, Surf. Interface Anal. **26**, 549 (1998).
[36] D. J. Morgan, Surf. Sci. Spectra **24**, 024004 (2017).
[37] P. Zhang, P. Richard, T. Qian, Y.-M. Xu, X. Dai, and H. Ding, Rev. Sci. Instrum. **82**, 043712 (2011).
[38] A. Savitzky and M. J. E. Golay, Anal. Chem. **36**, 1627–1639 (1964).
[39] P. E. Blöchl, Phys. Rev. B **50**, 17953 (1994).
[40] J. P. Perdew, K. Burke, and M. Ernzerhof, Phys. Rev. Lett. **77**, 3865 (1996).
[41] J. P. Perdew, J. A. Chevary, S. H. Vosko, K. A. Jackson, M. R. Pederson, D. J. Singh, and C. Fiolhais, Phys. Rev. B **46**, 6671 (1992).
[42] C. D. Langreth and M. J. Mehl, Phys. Rev. B **28**, 1809 (1983).
[43] G. Kresse, J. Non-Cryst. Solids **3093**, 222 (1995).

[44] V. Wang, N. Xu, J.-C. Liu, G. Tang, and W.-T. Geng, Comput. Phys. Commun. **267**, 108033 (2021).

[45] W.-T. Geng, Y.-C. Liu, N. Xu, G. Tang, Y. Kawazoe, and V. Wang, Nat. Protoc. **20**, 3143–3169 (2025).

[46] S. Duan and X. Xu, J. Phys. Chem. Lett. **14**, 6726–6735 (2023).

[47] J. Tersoff and D. R. Hamann, Phys. Rev. B **31**, 805 (1985).

[48] A. Marrazzo, S. Beck, E. R. Margine, N. Marzari, A. A. Mostofi, J. Qiao, I. Souza, S. S. Tsirkin, J. R. Yates, and G. Pizzi, Rev. Mod. Phys. **96**, 045008 (2024).

[49] G. Pizzi, V. Vitale, R. Arita, S. Blügel, F. Freimuth, G. Géranton, M. Gibertini, D. Gresch, C. Johnson, T. Koretsune, J. I. Ibañez-Azpiroz, H. Lee, J. M. Lihm, D. Marchand, A. Marrazzo, Y. Mokrousov, J. I. Mustafa, Y. Nohara, Y. Nomura, L. Paulatto, S. Poncé, T. Ponweiser, J. Qiao, F. Thöle, S. S. Tsirkin, M. Wierzbowska, N. Marzari, D. Vanderbilt, I. Souza, A. A. Mostofi, and J. R. Yates, J. Phys.: Condens. Matter **32**, 165902 (2020).

[50] S. S. Tsirkin, npj Comput. Mater. **7**, 33 (2021).

[51] J.-F. Qiao, J.-Q. Zhou, Z. Yuan, and W.-S. Zhao, Phys. Rev. B **98**, 214402 (2018).

[52] D. Gresch, G. Autès, O. V. Yazyev, M. Troyer, D. Vanderbilt, B. A. Bernevig, and A. A. Soluyanov, Phys. Rev. B **95**, 075146 (2017).

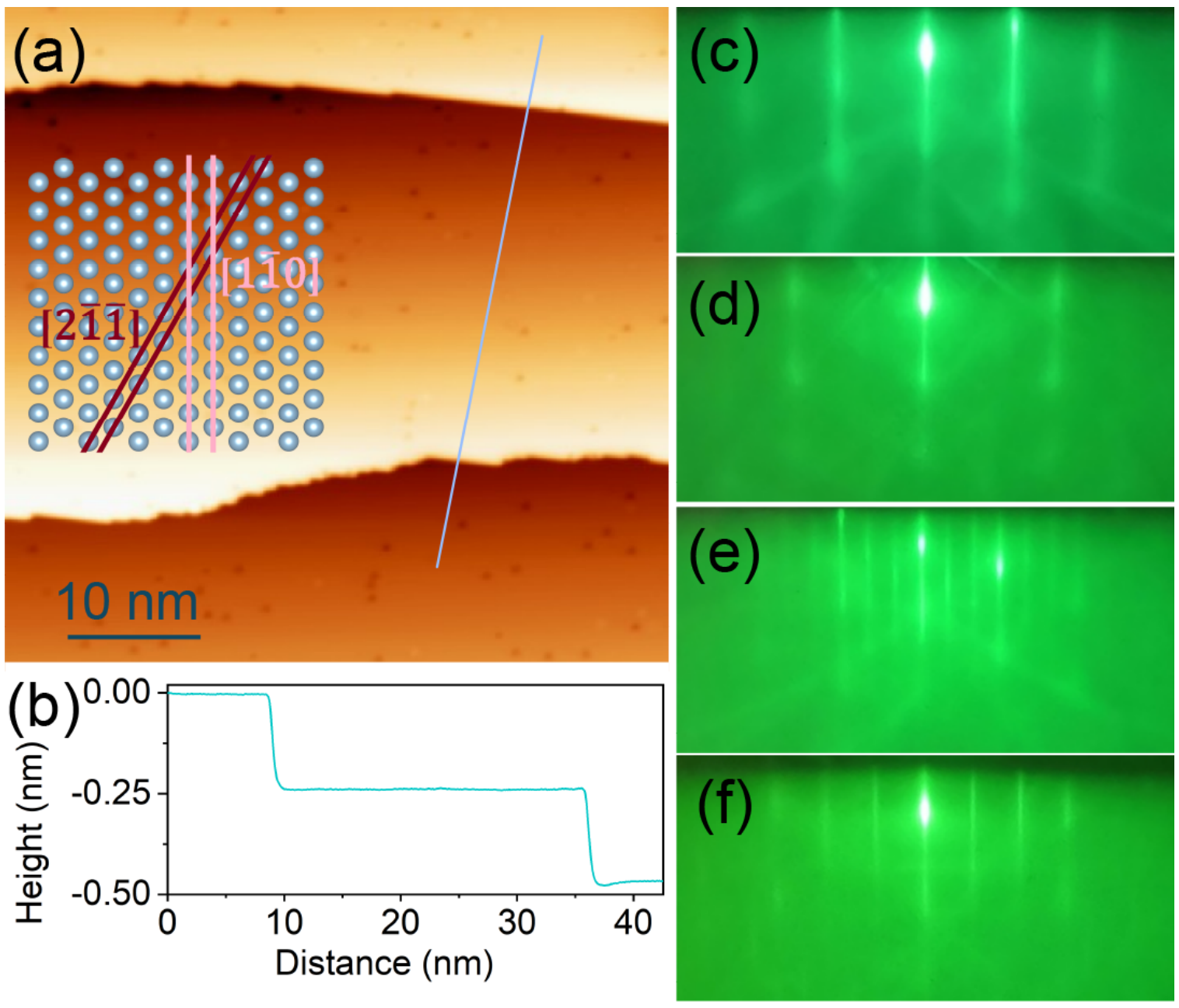

FIG. 1. (a) 50 nm ×50 nm large area STM topograph of bare Al(111), acquired at a sample bias of $V_b$ = +1 V and $I_t$ = 100 pA. The inset shows a schematic top view of the Al(111) surface lattice, with the [2$\bar{1}\bar{1}$] and [1$\bar{1}$0] crystallographic directions indicated. The oblique blue line denotes the path along which the line profile was extracted. (b) Apparent-height profile along the blue line in (a), showing two steps of approximately 0.24 nm, consistent with monatomic Al(111) steps. (c), (d) RHEED patterns of bare Al(111) recorded along the [1$\bar{1}$0] and [2$\bar{1}\bar{1}$] azimuths, respectively. (e), (f) Corresponding RHEED patterns obtained after Sb deposition along the same respective azimuths.

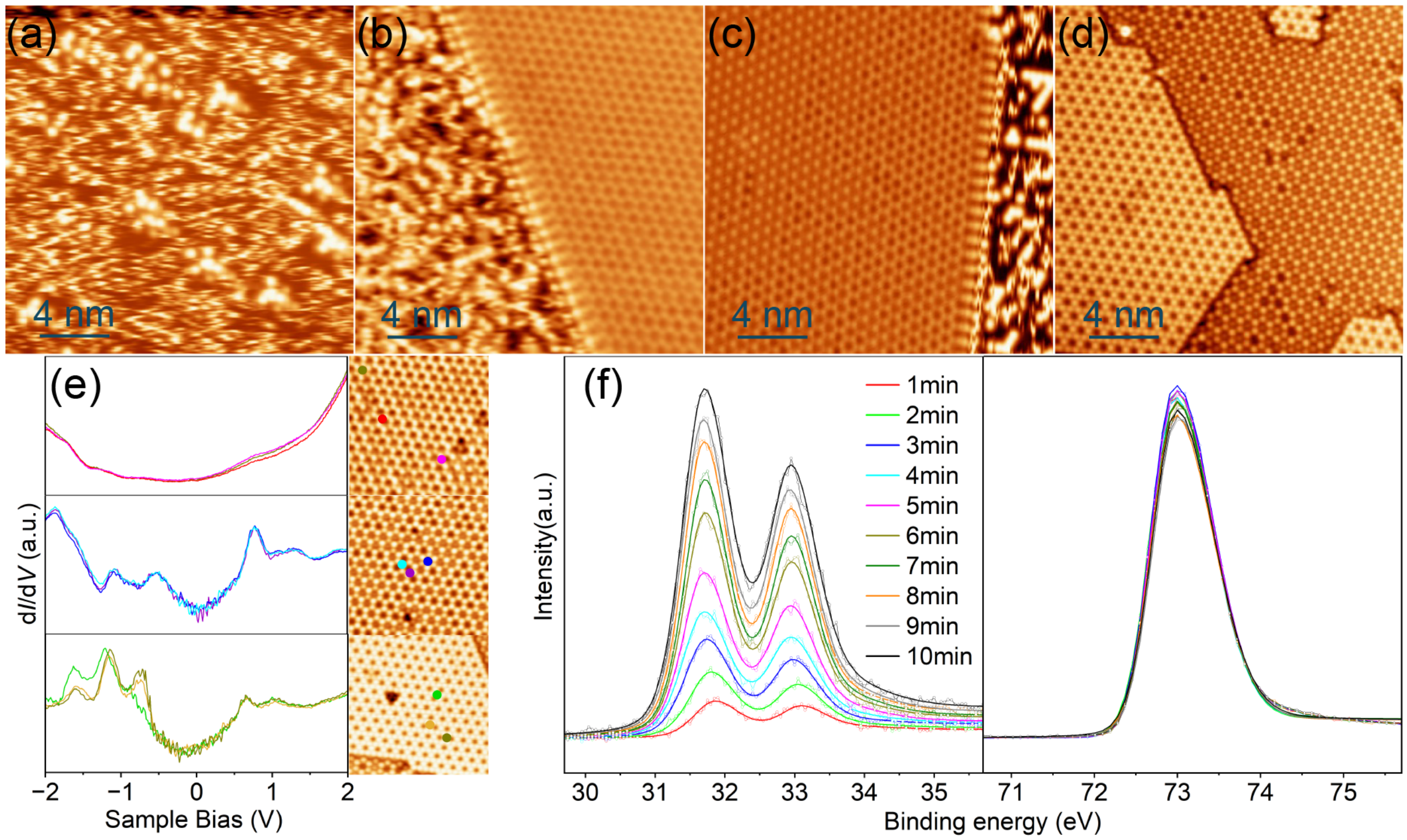

FIG. 2. (a)–(d) Representative constant-current STM topographs (20 nm × 20 nm) of Sb/Al(111) acquired after 2, 4, 6, and 8 min of Sb deposition, respectively. Tunneling parameters: (a) $V_b$=+0.5 V and $I_t$=500 pA; (b) $V_b$=+0.4 V and $I_t$=500 pA; (c) $V_b$=+1.0 V and $I_t$=100 pA; (d) $V_b$=−0.5 V and $I_t$=310 pA. (e) Differential conductance d$I$/d$V$ spectra acquired at three positions for each deposition time. The top, middle, and bottom sets correspond to 4, 6, and 8 min of Sb deposition, respectively; the colors match the marked acquisition positions in the adjacent STM topographs. (f) Sb 4$d$ (left) and Al 2$p$ (right) core level XPS spectra of Sb/Al(111) acquired after 1–10 min of deposition. Symbols and colored solid curves denote the experimental data and total fits, respectively.

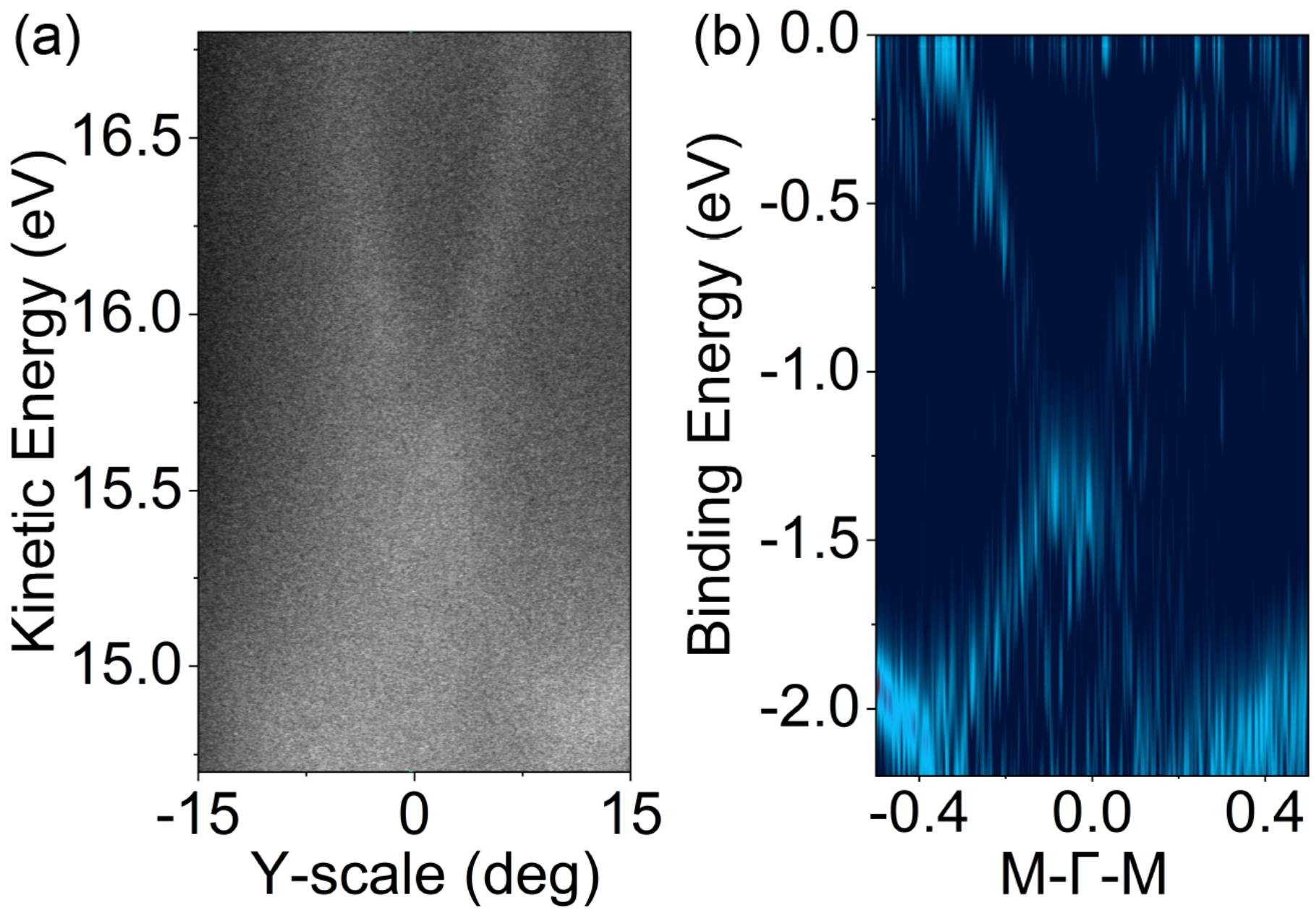

FIG. 3. (a) Angle-resolved photoemission intensity map as a function of kinetic energy and emission angle, acquired using hν = 21.218 eV. (b) Corresponding map of binding energy and momentum along M-Γ-M.

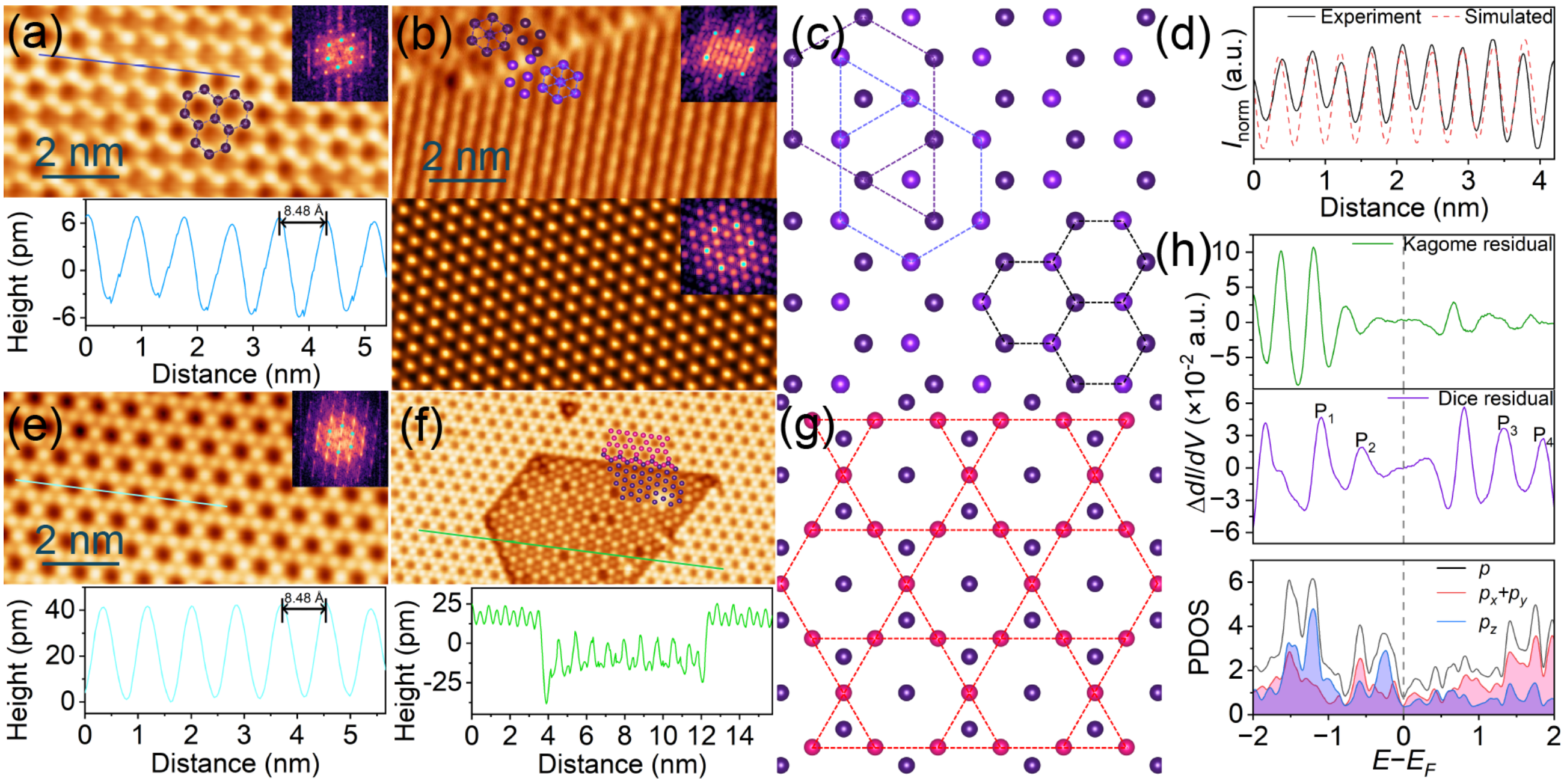

FIG. 4. (a) Constant current STM topograph of a 10 nm × 5 nm area, acquired at a sample bias of $V_b$=+0.4 V and a tunneling-current setpoint of $I_t$=500 pA. The inset and lower panel show the corresponding fast Fourier transform (FFT) and apparent height profile along the blue line. (b) Experimental constant current STM image of a 10 nm × 5 nm area of the dice phase formed (upper), acquired at a sample bias of $V_b$=+0.5 V and a tunneling-current setpoint of $I_t$=600 pA, and the corresponding simulated image (lower); the insets show their respective FFTs. (c) Top view of multiple repeated dice lattice with the upper (darker purple atoms) and lower (lighter purple atoms) layers stacked in an AB registry. (d) Normalized strip-averaged profiles from (b). (e), (f) Constant current STM topographs showing 10 nm × 5 nm and 20 nm × 10 nm areas, respectively. Both images were acquired at a sample bias of $V_b$=−0.5 V, with tunneling-current setpoints of $I_t$=310 pA for (e) and 700 pA for (f). The inset in (e) shows the corresponding FFT; the lower panels show apparent-height profiles along the cyan and green lines. (g) Schematic top view of the kagome-like electronic texture, with red dashed lines outlining the kagome network. (h) STS residuals for the kagome-like and dice regions and the calculated $p$ orbital PDOS of the dice lattice model (lower panel).

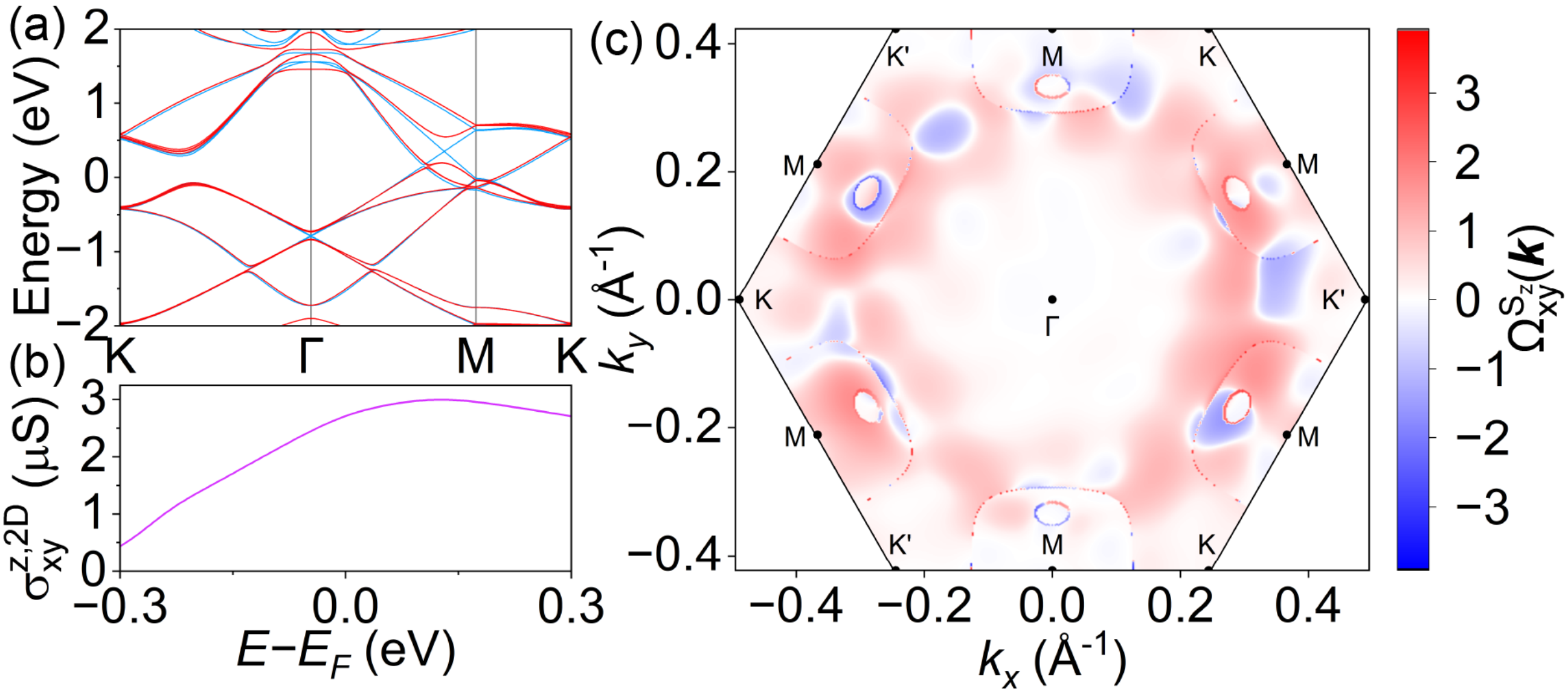

FIG. 5. (a) DFT calculated electronic band structures of β-Sb with (red) and without (blue) SOC. (b) Calculated intrinsic 2D spin Hall conductivity $\sigma_{xy}^{z,2D}$ of β-Sb as a function of energy relative to the Fermi level. (c) Momentum space distribution of the $z$-polarized spin Berry curvature $\Omega_{xy}^{S_z}(\boldsymbol{k})$ of β-Sb over the first Brillouin zone at $E_F$.